\documentclass[9pt,twocolumn,aps,prl,superscriptaddress,amssymb,nofootinbib,nobibnotes]{revtex4-1} 
\usepackage{multirow}
\usepackage{comment}
\usepackage[utf8]{inputenc}
\usepackage{amsmath,amsfonts,amssymb,amsthm,color}
\usepackage{stmaryrd}
\usepackage{tikz}
\usepackage{graphicx}
\usepackage[sort&compress]{natbib}
\usepackage{xcolor}
\usepackage{multirow}

\renewcommand{\div}{\rm{div}}

\begin{document}

\title{Influence of lung physical properties on its flow--volume curves using a detailed multi-scale mathematical model of the lung} 

\author{Riccardo Di Dio}
\affiliation{Université Côte d'Azur, CNRS, LJAD, Vader Center, Nice, France}
\affiliation{Université Côte d'Azur, INRIA, Sophia Antipolis, France.}
\author{Michaël Brunengo}
\affiliation{Université Côte d'Azur, CNRS, LJAD, Vader Center, Nice, France}
\affiliation{Respinnovation SAS, Sophia Antipolis, France.}
\author{Benjamin Mauroy}
\email{benjamin.mauroy@univ-cotedazur.fr}
\affiliation{Université Côte d'Azur, CNRS, LJAD, Vader Center, Nice, France}

\date{\today}

\begin{abstract}
We develop a mathematical model of the lung that can estimate independently the air flows and pressures in the upper bronchi. 
It accounts for the lung multi-scale properties and for the air-tissue interactions.
The model equations are solved using the Discrete Fourier Transform, which allows quasi instantaneous solving, in the limit of the model hypotheses.
With this model, we explore how the air flow--volume curves are affected by airways obstruction or by change in lung compliance.
Our work suggests that a fine analysis of the flow--volume curves might bring information about the inner phenomena occurring in the lung.
\end{abstract}


\maketitle

\section{Introduction}

The human lung is a complex organ whose role is to bring oxygen from ambiant air to blood and to remove carbon dioxide from the body \cite{weibel_pathway_1984}.
In order to perform its function, the lung has a specific tree-shaped structure called the bronchial tree, which allows to bring the inhaled air in contact with blood using a large exchange surface folded into the thorax.
Air flows into airways -- the bronchi -- that bifurcate regularly into smaller airways. 
The human bronchial tree consists in about $17$ successive levels of bifurcating and size-decreasing airways.
Each terminal airway of the bronchial tree is connected to the acini in which the exchanges between air and blood occur. 
A schematic representation of the lung is shown in Figure \ref{tree}.
In order to localize the airways in the tree, we use the generation index: the generation of an airway corresponds to the number of airways on the path linking that airway and the trachea.
The geometry of the airways and the compliance of the lung play an important role on determining the amount of air flowing through the lung.
Actually, it has been shown that the geometry and the control of ventilation are optimized to minimize the energy spent for ventilating the lung \cite{noel_origin_2022}.
However, many prevalent diseases are affecting the geometry of the airways and the compliance of the lung, such as asthma, bronchiolitis, COPD, cystic fibrosis, etc.
The airways lumen areas are reduced either by constrictions (inflammation, smooth muscles contractions, etc.) or by obstruction, typically from an excess of bronchial mucus.
The consequence is an increase of the hydrodynamic resistance of the lung.
Then, the airflow in the patient lung can be decreased, inducing hypoxia, or the energy spent for ventilation is increased, potentially inducing exertion of the patient.

The lung airflows redistribution and decrease during pathologies is not well understood as of today.
The lung is a complex organ in which experimental studies are difficult to perform.
Actually, routine medical exploration of the lung is often based on mouth airflow analysis and most particularly on flow-volume curves.
However, no precise information about the inner processes occurring in the lung during pathologies is easily accessible.
Mathematical and numerical models are adapted to reach otherwise unreachable information on the airflows distribution.
However, even with these models, due to the complexity of the organ, many studies focus on lung subregions, either distal (deep) or proximal (large airways), as modeling the whole lung is computationally very costly.

In this work, we propose a new approach based on a low cost fluid-structure interaction model of the whole lung.
This model includes the interactions between the airflows in the airways and the parenchyma.
With this model, we can explore how a change in the lung characteristics affects the flow--volume curves.

\section{Methodology}

We develop a mathematical model of the lung that couples a model of airflows inside the airways tree and a model of elastic deformation of the parenchyma, inspired from the model developed in \cite{brunengo_optimal_2021}.\newline

\begin{figure*}[t!]
%\begin{center}
\centering
A \includegraphics[width=9.5cm]{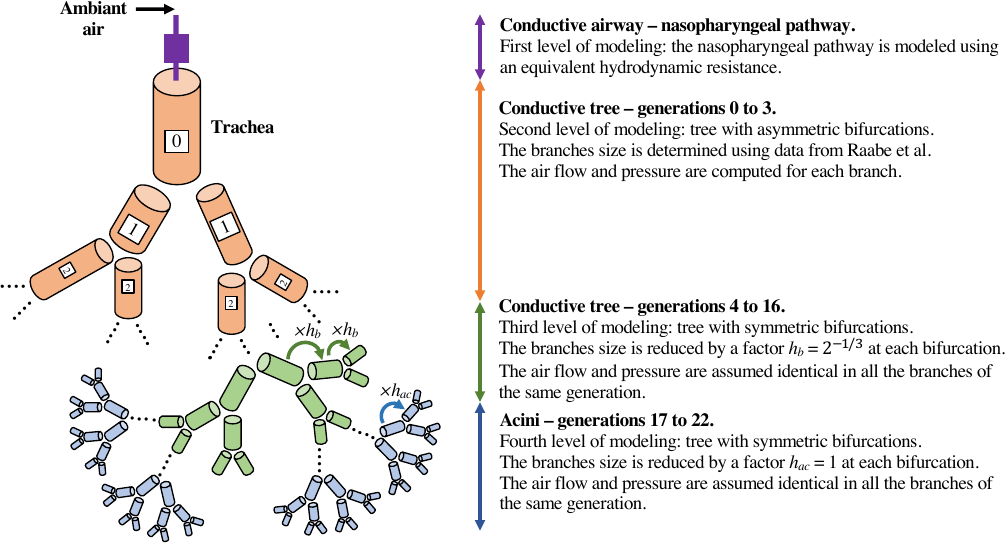}
\hspace{0.1cm}
B \includegraphics[width=6.5cm]{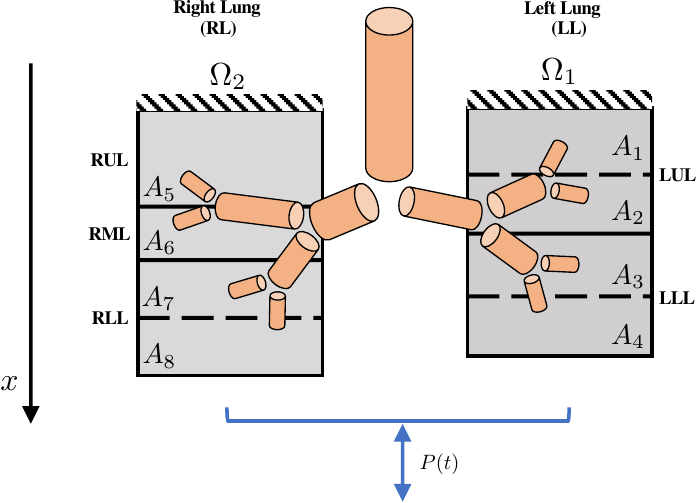}
\caption{{\bf A}: Schematic representation of the lung and of the model used in this work. 
The bronchial tree is divided into four levels of modelling. 
{\bf B}: The parenchyma is modeled as two elastic mediums $\Omega_1$ and $\Omega_2$ corresponding to the right and left lung.
Each domain $\Omega_i$ is divided into four regions $\left(\Omega_{2(i-1) + j}\right)_{j=1,\dots,4}$ adapted to the lung lobes or sub-lobes in terms of localisation and volumes.
Full lines divide the lobes, while dashed lines divide sub-lobes (RUL: right upper lobe; RML: right medium lobe; RLL: right lower lobe; LUL: left upper lobe; LLL: left lower lobe).
Each regions is fed by a terminal branch form the second level of modeling of the bronchial tree.
}
\label{tree}
\end{figure*}

{\bf Model of the bronchial tree and acini.}
The model of the airway tree consists in an assembly of four levels of modeling, see Figure \ref{tree}. 
Each level consists in tree-like structure(s) with cylindrical rigid airways.
The airflows are determined using pressure-flow relationships based on the hydrodynamic resistances of the airways and on equivalent resistances of the different levels of modeling, see \cite{brunengo_optimal_2021}.

The first level corresponds to the nasopharyngeal pathway, which is modeled as a hydrodynamic resistance of $48 750$ Pa.m$^{-3}$.s \cite{martonen_flow_2002}.
The second level corresponds to the four first generations, from $0$ to $3$.
Each airway size in this level comes from Raabe et al. data \cite{raabe_tracheobronchial_1976}.
The trachea is modeled as the single airway from generation $0$ and is connected to the nasopharyngeal pathway.
The hydrodynamic resistance of the airways are corrected with an ad-hoc factor that accounts for inertial effects in the airways. 
It is calibrated so that the hydrodynamic resistance of the whole model of the lung fits the physiological value at rest \cite{brunengo_optimal_2021}.  
The pressures-flows equations of the airways are solved using a resistance matrix, see \cite{brunengo_optimal_2021}.
The third level corresponds to the generations five to sixteen.
These levels of the bronchial tree are modeled using a symmetric bifurcating tree as Weibel's model A \cite{weibel_pathway_1984}. 
At each bifurcation, the size of the airways decreases by a factor $h_b = (\frac12)^\frac13 \simeq 0.79$.
Each terminal airways of the second level is connected to two trees of the third level.
The influence on the airflows of the resulting tree is represented by the equivalent resistance of the tree.
The fourth level of modeling corresponds to the acini and is similar to the third level, except that the reduction factor at bifurcations is now $h_{\rm{ac}} = 1$ \cite{noel_origin_2022}.
Each terminal airways of the third level is connected to two trees of the fourth level.

All the levels are coupled together using pressure-flow relationships and airflows conservation.
In the second level, the air pressures and flows depend on each airway.
Since the airway tree models of the third and fourth levels are based on symmetric bifurcations, the air pressures and flows in a subtree of these levels depend only on the generation index and on the second level terminal branch to which the subtree is connected.
Given a distribution of airflows $F=(F_k)_{k=1,\dots,8}$ in the terminal airways of the second level, this model allows to compute the air pressure $p=(p_k)_{k=1,\dots,8}$ inside the terminal branches of the acini using the resistance matrix $\mathcal{R}$ resulting from our model: $p = \mathcal{R} F$.\newline

{\bf Model of the parenchyma.}
We consider that the parenchyma is decomposed into two domains $\Omega_1$ and $\Omega_2$ corresponding to the left lung (LL) and the right lung (RL).
Each of the domain $\Omega_i$ is divided into four regions $\left(A_{2(i-1) + j}\right)_{j=1,\dots,4}$, for a total of eight regions for the whole lung.
Each region is considered as a lung lobe or sub-lobe and is connected to one of the terminal branch of the second level of modeling. 
The volume of each region is then determined by identifying which lung lobe the corresponding terminal branch is feeding in Raabe et al. data.
Lung lobes volume are gathered from \cite{yamada_differences_2020}.
The volume of sub-lobes are assumed to be half the volume of the corresponding lobe.

Both the left and right lungs are assumed to deform only along the vertical axis $x$, see Figure \ref{tree}B and to have a section perpendicular to that axis of $S_1$ and $S_2$.

%\begin{wrapfigure}{l}{0.6\linewidth}%[t!]
\begin{figure*}[ht!]
%\centering
%\begin{center}
%A \includegraphics[width=4cm]{VentSignal-crop.pdf}
\hspace{-1cm}\includegraphics[width=15cm]{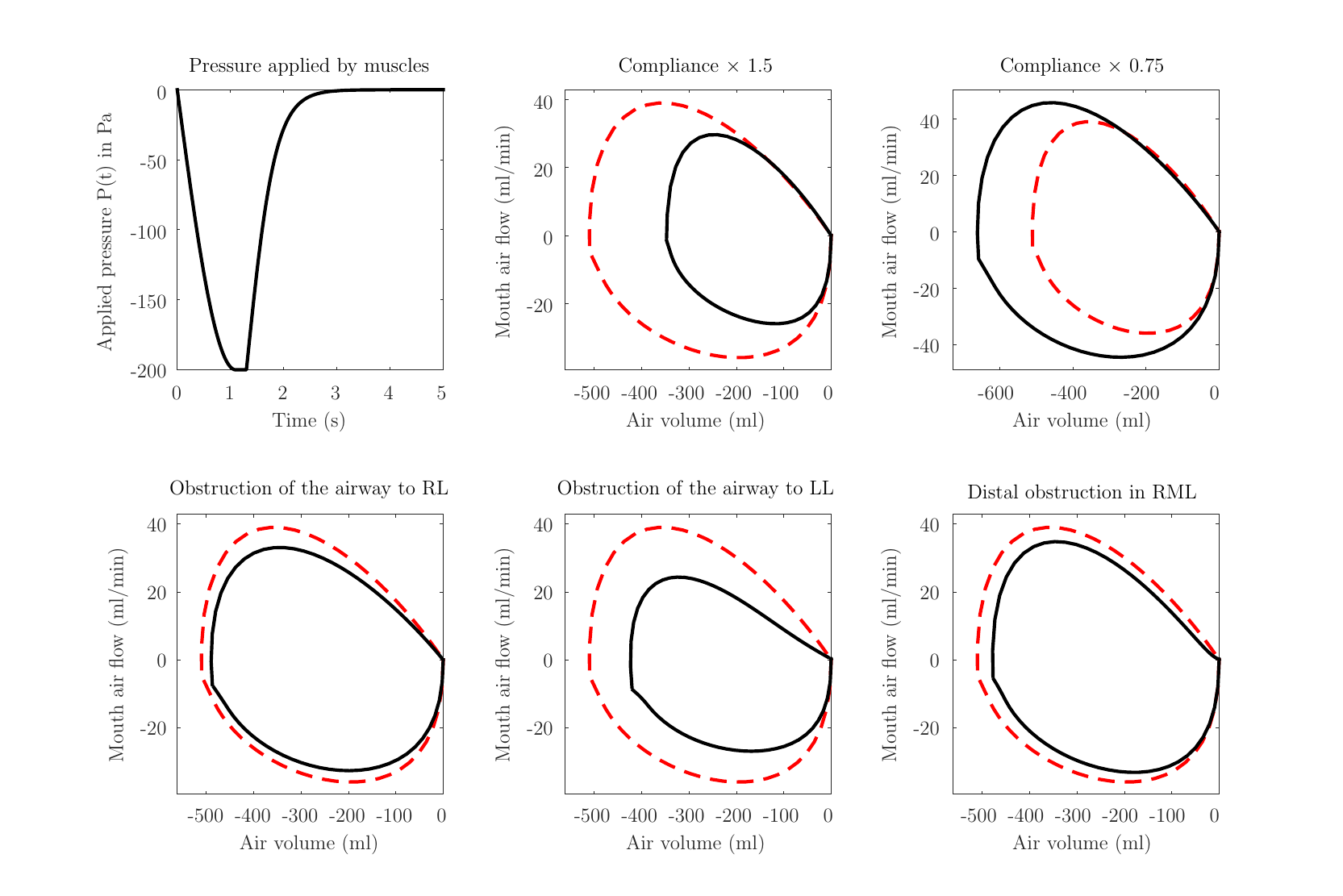}
\caption{Flow--volume curves predicted by our model for a set of scenarios.
Not only are the amplitudes of pressures and volumes affected, but also the shapes of the curves.}
\label{results}
%\end{center}
%\end{wrapfigure}
\end{figure*}

Denoting $\lambda = 1800$ Pa and $\mu = 450$ Pa the lamé coefficient of the lung parenchyma \cite{brunengo_optimal_2021}, the stress--strain relationship in $A_{2(i-1)+j}$ is $\sigma(u_i) = \lambda \rm{Tr}(\epsilon(u_i)) \rm{I} + 2 \mu \epsilon(u_i) - p_{2(i-1)+j} \rm{I}$  \cite{brunengo_optimal_2021} where $p = \mathcal{R} F$. 
The flow $F_k$ in $A_k$ is the rate of change of the volume of $A_k$ assuming small deformations, i.e. $F_k = - S_i \int_{A_k} \div\left( \frac{\partial u_i}{\partial t} \right) dx$ with $i = \lfloor (k-1)/4 \rfloor + 1$.
We assume that the $A_i$ are fixed in $x=0$ and submitted to the same periodic pressure $P(t)$ mimicking the ventilation muscle action in $x=L_i$ \cite{brunengo_optimal_2021}, see Figure \ref{tree}B.\newline

{\bf Model resolution.}
The model resolution is based on two steps.
First, since the equations are unidimensional in space, we could derive an analytical solution for any applied pressure $P(t)$ of the form $P(t) = A \cos(\omega t) + B \sin(\omega t)$ \cite{brunengo_optimal_2021}.
Second, the equations of the model are linear relatively to the pressure signal $P(t)$.
Thus, we use a sampled version of $P(t)$, typically with $128$ time points for a ventilation cycle.
We decompose the sampled $P(t)$ with the Discrete Fourier Transform (DFT) using the Fast Fourier Transform algorithm.
Then, we can solve analytically the model equation for each frequency component of the DFT, which is in the form $A \cos(\omega t) + B \sin(\omega t)$.
This step can also benefit of parallel computing, as each component can be computed separately.
Indeed, if the number of generations of the second level is $n$, then the matrix $\mathcal{R}$ is a full matrix of size $2^{n-1} \times 2^{n-1}$. 
This matrix has to be inverted in order to solve the model equations.
Finally, the last step is to reconstruct the full solution by adding the contribution of each frequency.

Hence, the computing cost for solving the model equations remains in general very low if the number of generations of the second level of modeling is not too large, as in the study made in this work ($n=4$). 
We solved the system using the software {\it GNU Octave}.

\section{Results and conclusions}

Our model is able to explore a wide range of scenarios, either in terms of change in compliance or in terms of obstruction of the airways.
These scenarios should remain close to normal ventilation as our model assumes linear elasticity and has been calibrated to mimic normal ventilation.

In order to mimic routine clinical functional exploration, we tested how the flow--volume curves were affected by the scenarios.
Our results suggest that not only are the amplitudes of flows and volumes affected, but also the shapes of the flow--volume curves.
Several scenarios are plotted in Figure \ref{results}, where positive air flows indicate expiration and negative air flows indicate inspiration.
In our model, inspiration is active, i.e. driven by muscles, while expiration is passive, i.e. driven by the elastic recoil of the parenchyma.
As a consequence, inspiration and expiration are not symmetrical in time, as shown in Figure \ref{results}.

This induces that inspiration and expiration do not respond identically to a change in the model parameters.
For example, a change in compliance does not affect much the flow--volume curve at the end of expiration and start of inspiration, while an obstruction does.
Similarly, the obstruction of an airway of generation $2$ of $50 \%$ of its diameter leads to a different response on the flow--volume curve depending on whether it is the airway feeding the left or the right lung.
The obstruction of the airway feeding the left lung induces a strong change in the shape of the flow--volume curve during expiration.
This airway is smaller than the one feeding the right lung, hence it is more sensible to an obstruction.
More generally, the two airways have different geometries and feed regions with different sizes, hence the system responds differently to an obstruction of one of them.

Our results also suggest that the expiration, the passive phase of lung ventilation, ought to be a better indicator of a pathology than inspiration, which is actively driven. 
Indeed, the inspiration parts of the flow--volume curves are all quite similar whatever the scenario.

Hence, flow--volumes curves at expiration might be used to get information about the inner phenomena occurring in the lung and most particularly about the airways obstructions.
Further studies should be made to characterize more precisely how the flow--volume curves are affected by a change of parameters.
This work represents however a first step toward this characterization and should be considered as a proof of concept.
In our model, several biophysical phenomena have been neglected, in particular the non-linear response of the air flows and of the mechanics of the parenchyma.
These phenomena are known to affect flow--volume curves, mainly during forced expiration and they might help to get more information.

It is important to notice that our model predictions are made in an ideal framework, which allows to get neat data. 
But the differences observed between the flow--volume curves might be too small or noisy to be observed in real functional exploration data.
Hence, once the characterization of the flow--volume curves have been performed, it would be important to test the applicability of the characterization to experimental flow--volume curves.

\section{Acknowledgements}
This project has received funding from the European Union’s Horizon 2020 research and innovation programme under the Marie Curie grant agreement No 847581 and is co-funded by the Region SUD Provence-Alpes-Cote d’Azur and IDEX UCA JEDI.
\begin{figure}[h!]
\includegraphics[width=8cm]{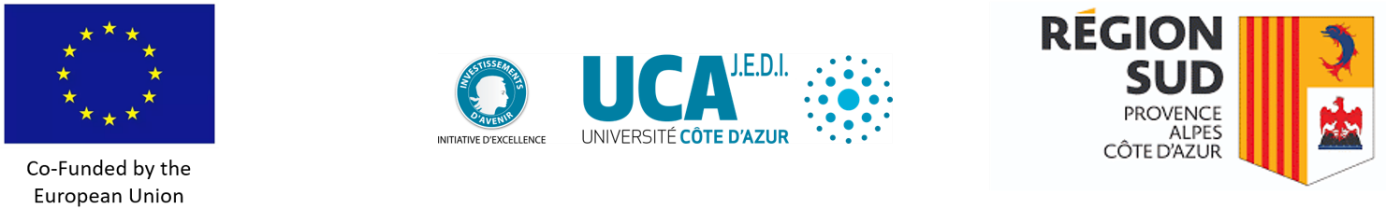}
\end{figure}

%\bibliographystyle{abbrv}
%\setlength{\parskip}{0pt}
%\bibliography{Biblio}

\end{document}